# Gauge parameter dependence in gauge theories

E. Kraus[a] and K. Sibold[b]

[a]Institut f. Theoretische Physik, Univ. Bern, Sidlerstr. 5, CH-3012 Bern, Switzerland

[b]Max-Planck-Inst. f. Physik, Werner-Heisenberg-Institut,
Föhringer Ring 6, D-80805 München, Germany

Dependence on the gauge parameters is an important issue in gauge theories: physical quantities have to be independent. Extending BRS transformations by variation of the gauge parameter into a Grassmann variable one can control gauge parameter dependence algebraically. As application we discuss the anomaly coefficient in the Slavnov-Taylor identity, $S$-matrix elements, the vector two-point-function and the coefficients of renormalization group and Callan-Symanzik equation.

## 1. The problem and its solution

To begin with we write down the invariant action of a Yang-Mills theory with simple gauge group

$$\Gamma_{inv} = -\frac{1}{4g^2} \int Tr(\partial_\mu A_\nu - \partial_\nu A_\mu + i[A_\mu, A_\nu])^2 + \Gamma_{\text{matter}}. \quad (1)$$

(We use matrix notation and do not display explicitly the matter part.) In perturbation theory one has to invert the bilinear part of the vector fields which requires adding a gauge fixing term. The most common one is

$$\Gamma_{\text{g.f.}} = \int \frac{-1}{2\xi} Tr(\partial A)^2 \quad (2)$$

It is obviously important to control the dependence of the theory on the parameter $\xi$, because gauge fixing introduces unphysical degrees of freedom into the theory, namely the scalar and longitudinal components of the vector field. In (massless) Yang-Mills theory they interact, spoil unitarity and thus have to be controlled. Any quantity will therefore qualify as "physical" only if it is independent of $\xi$. (Whether, conversely, a $\xi$-independent quantity is physical, is a subtle question. S.b.) $\xi$-dependence for the gauge fixing (2) has been studied [1], but it requires heavy tools like the Wilson expansion.

In order to prove renormalizability of gauge theories and unitarity of the physical S-matrix to all orders of perturbation theory in a scheme independent way, it is essential to have the BRS transformations in a form which can be simply controlled algebraically, i.e. they have to be nilpotent in their action on all fields including the Faddeev-Popov ($\phi\pi$)-fields $c$ and $\bar c$. For these purposes one couples the gauge condition linearly to an auxiliary field $B$ of dimension 2, which acts as a Langrange multiplier:

$$\Gamma_{\text{g.f.}} = \int Tr(\tfrac{1}{2}\xi B^2 + B\partial A) \quad (3)$$

Eliminating $B$ with its equ. of motion yields (2). In fact this term can be completed to a BRS invariant

$$\Gamma_{\text{g.f.}} + \Gamma_{\phi\pi} = \int Tr(\tfrac{1}{2}\xi B^2 + B\partial A - \bar c \partial(\partial c + i[c, A])) \quad (4)$$

with nilpotent transformation law for all fields

$$s\bar c = B \qquad sA = \partial c + i[c, A] \quad (5)$$
$$sB = 0 \qquad sc = icc$$
$$s^2\phi = 0 \qquad \text{for } \phi = A, \bar c, c, \ldots \quad (6)$$

Moreover one can write this invariant as a variation

$$\Gamma_{\text{g.f.}} + \Gamma_{\phi\pi} = s \int Tr(\tfrac{1}{2}\xi \bar c B + \bar c \partial A) \quad (7)$$

This fact is important for interpretation: terms which are BRS variations do not contribute between physical states (s. 2.2 below). Especially the auxiliary field $B$ drops out from the physical S-matrix of the theory, and is – as mentioned above – introduced because of technical reasons making the algebraic structure of the Green functions more transparent.

Along the same lines we observe that

$$\partial_\xi \Gamma_{\text{g.f.}} = \int \tfrac{1}{2} Tr(B^2) = s \int \tfrac{1}{2} Tr(\bar{c}B) \tag{8}$$

too, is a variation. By permitting $\xi$ to vary into a Grassmann variable $\chi$ and completing the transformations to be nilpotent again

$$s\xi = \chi \quad s\chi = 0 \tag{9}$$

we maintain extended BRS transformations. Consequently the BRS-invariant associated with the gauge fixing (3) has to be enlarged by a $\chi$-dependent term

$$\Gamma_{\text{g.f.}} + \Gamma_{\phi\pi} = \int Tr(\tfrac{1}{2}\xi B^2 + \tfrac{1}{2}\chi\bar{c}B + B\partial A \\ - \bar{c}\partial(\partial c + i[c,A])) \tag{10}$$

For all concrete calculations one can return to the original form of BRS transformations and the action by taking $\chi = 0$, but as we show below one has constructed a tool to control the dependence of the Green functions on the gauge parameter in a completely algebraic way.

These considerations can be continued to all orders of perturbation theory. If one does not want to stick to a specific renormalization scheme it is best to introduce external fields coupled to the non-linear BRS variations.

$$\Gamma_{\text{ext.f.}} = \int Tr(\rho^\mu sA_\mu + \sigma sc) + \ldots \tag{11}$$

(the dots stand for matter contributions). BRS invariance is then governed by the extended Slavnov-Taylor identity (ST)

$$\mathcal{S}(\Gamma) \equiv \int Tr\left(\frac{\delta\Gamma}{\delta\rho}\frac{\delta\Gamma}{\delta A} + \frac{\delta\Gamma}{\delta\sigma}\frac{\delta\Gamma}{\delta c} + B\frac{\delta\Gamma}{\delta\bar{c}}\right) \\ + \chi\partial_\xi\Gamma = 0 \tag{12}$$

(matter contributions are suppressed).

$\partial_\xi\Gamma$ is now a BRS variation in the functional sense

$$\partial_\xi\Gamma = -\partial_\chi \int Tr\left(\frac{\delta\Gamma}{\delta\rho}\frac{\delta\Gamma}{\delta A} + \frac{\delta\Gamma}{\delta\sigma}\frac{\delta\Gamma}{\delta c} + B\frac{\delta\Gamma}{\delta\bar{c}}\right)$$

$$= \mathcal{S}_\Gamma(\partial_\chi\Gamma) \tag{13}$$

$$\mathcal{S}_\Gamma \equiv \int Tr\left(\frac{\delta\Gamma}{\delta\rho}\frac{\delta}{\delta A} + \frac{\delta\Gamma}{\delta A}\frac{\delta}{\delta\rho} + \frac{\delta\Gamma}{\delta\sigma}\frac{\delta}{\delta c} + \frac{\delta\Gamma}{\delta c}\frac{\delta}{\delta\sigma}\right. \\ \left. + B\frac{\delta}{\delta\bar{c}}\right) \tag{14}$$

Controlling $\xi$-dependence thus relies on establishing (12). Since this amounts to solving only an algebraic problem [2,3] we speak of "algebraic" control of gauge parameter dependence.

## 2. Applications

### 2.1. Anomaly coefficient

For the proof of (12) one can take over the technique of [1]. One permits all possible breaking terms on the r.h.s. of (12) and uses consistency conditions implied by the symmetry algebra to constrain them. Somewhat more specifically one has

$$\mathcal{S}(\Gamma) = \hbar\Delta \tag{15}$$

in the one-loop approximation, if one starts with a BRS invariant classical action. Here $\Delta$ is a sum over all local field monomials of dimension four and $\phi\pi$-charge $+1$. As constraint from the algebra arises

$$s\Delta = 0, \tag{16}$$

where not only the fields vary according to (5), but also $\xi, \chi$ according to (9). I.e. $\xi$-dependence in the coefficients is taken care of by $s$. It turns out that all breaking terms can be counterbalanced by counterterms to the classical action, but one potential obstruction candidate survives

$$\mathcal{S}(\Gamma) = r\mathcal{A} + o(\hbar r\mathcal{A}) \tag{17}$$

$$\mathcal{A} = \varepsilon_{\mu\nu\rho\sigma}\int Tr(c\partial^\mu(\partial^\nu A^\rho A^\sigma \\ -\tfrac{1}{2}A^\nu A^\rho A^\sigma)) \tag{18}$$

– the chiral anomaly. The coefficient $r$ has to be $\xi$-independent

$$\partial_\xi r = 0. \tag{19}$$

This is a crucial property since non-vanishing $r$ would lead to violation of unitarity and in this sense $r$ is a physical quantity. It can be shown that it *must* start at one-loop (the non-renormalization theorem of [4]) and hence is forced to vanish to all orders by suitably arranging the content of the fermion representation $T$:

$$r \sim d^{ijk} Tr T^i \{T^j, T^k\} \stackrel{!}{=} 0 \qquad (20)$$

($d^{ijk}$ is the totally symmetric tensor of the gauge group.)

By recursion one can then show that (12) holds to all orders.

## 2.2. S-matrix elements

Rewriting (13) for the generating functional of general Green functions

$$\partial_\xi Z = \mathcal{S}(\partial_\chi Z) \qquad (21)$$

$$\mathcal{S} \equiv \int \left( Tr(j_A \frac{\delta}{\delta\rho} + j_c \frac{\delta}{\delta\sigma} + j_{\bar{c}} \frac{\delta}{\delta j_B}) + j_\phi \frac{\delta}{\delta Y_\phi} \right)$$

(here $j_\phi \delta/\delta y_\phi$ summarizes the matter contributions) we see, why $S$-matrix elements in the physical subspace are $\xi$-independent (in those cases where the $S$-matrix exists). For obtaining them one differentiates only with respect to $j_A$ and $j_\phi$ and projects then on the physical shell. But $\delta/\delta\rho$ and $\delta/\delta Y_\phi$ can generate one-particle-poles at most in the ghost sector. Hence the projection yields zero.

## 2.3. Vector two-point-function

In order to find out how the vector two-point-function depends on $\xi$, we test (13) twice with respect to $A$. Using $\phi\pi$-charge conservation we arrive immediately at

$$\partial_\xi \Gamma_{\mu\nu} = \int Tr \left( -\partial_\chi \frac{\delta^2 \Gamma}{\delta A^\mu \delta\rho} \frac{\delta^2 \Gamma}{\delta A \delta A^\nu} + (\mu \leftrightarrow \nu) \right) (22)$$

Since in the classical approximation there is no term $\int \chi A \rho$ in the action, (22) yields

$$\partial_\xi \Gamma^{(0)}_{\mu\nu} = 0 \qquad (23)$$

– in accordance with the fact, that $\xi$ appears only in front of the bilinear $B$-term. In the one-loop approximation (22) reads in momentum space

$$\partial_\xi \Gamma^{(1)}_{\mu\nu}(p) = \partial_\chi \frac{\delta^2 \Gamma^{(1)}}{\delta A^\mu(p) \delta\rho^\lambda(-p)} (-\eta^\lambda{}_\nu p^2 + p^\lambda p_\nu) +$$
$$(\mu \leftrightarrow \nu)$$
$$\equiv \gamma^{(1)}_{\mu\nu}(\xi, p) \qquad (24)$$

where the vertex function $\partial_\chi \Gamma^{(1)}_{A\rho}$ is given by two diagrams. They contain the vertex $\chi\bar{c}B$ and thus the mixed propagator $< T(A_\mu B) >$. Since they contain at most once a triple vector vertex and are only logarithmically divergent, they are much simpler than those diagrams making up the two-point function $\Gamma^{(1)}_{\mu\nu}$.

The integral $\int_{\xi_o}^{\xi} \gamma^{(1)}_{\mu\nu}$ therefore governs the complete $\xi$-dependence of the two-point-function $\Gamma^{(1)}_{\mu\nu}$ and can serve, e.g. the purpose of extending a calculation of the latter in any convenient gauge – defined by a gauge parameter $\xi_o$ – to a gauge with parameter $\xi$:

$$\hat{\Gamma}^{(1)}_{\mu\nu}(\xi, p) = \Gamma^{(1)}_{\mu\nu}(\xi_o, p) - \int_\xi^{\xi_o} \gamma^{(1)}_{\mu\nu}(\zeta, p) d\zeta \qquad (25)$$

(As long as the gauge invariance is not spontaneously broken the $\xi$-dependence is polynomial, hence the integration trivial.) If the two-point-function were a candidate for an observable (i.e. a physical quantity) this kind of redefinition would render it gauge-independent. One would then have to understand only the role of $\xi_o$.

It is clear that this procedure is not restricted to one-loop. Abbreviating the r.h.s. of (22) again by $\gamma^{(n)}_{\mu\nu}(\xi, p)$ one can define also at $n$-th order

$$\hat{\Gamma}^{(n)}_{\mu\nu}(\xi, p) = \Gamma^{(n)}_{\mu\nu}(\xi_o, p) - \int_\xi^{\xi_o} \gamma^{(n)}_{\mu\nu}(\zeta, p) d\zeta \qquad (26)$$

and connect in this way a gauge $\xi_o$ with a gauge $\xi$. Again the diagrams contributing to $\gamma^{(n)}_{\mu\nu}$ will be much simpler than those contributing to $\Gamma^{(n)}_{\mu\nu}$ and the latter has to be calculated only in a convenient gauge. In an analogous manner one can treat also vectorial three- and four-point functions.

This construction should be compared with the pinch-technique [5] and the analogous results presented in these proceedings by A. Denner.

## 2.4. $\beta$-functions
### 2.4.1. Rigid gauge invariance maintained

It is well known that the $\beta$-function associated with the Callan-Symanzik equ. (CS) appears in the scaling of $S$-matrix elements, whereas anomalous dimensions drop out [6]. In this sense $\beta$-functions are physical quantities and should be constructed as being $\xi$-independent, whereas for $\gamma$'s no such requirement holds. In refs.[2,3] it has been shown how this can be done for pure Yang-Mills theories in linear and nonlinear gauges: one constructs the CS- and renormalization group (RG)-differential operators such that they have definite symmetry properties with respect to extended BRS. It then turns out that indeed the $\beta$-function can be constructed as being $\xi$-independent whereas the $\gamma$'s in general depend on $\xi$. The same construction works in the presence of matter fields as long as the rigid gauge invariance is not spontaneously broken. In formulas these results read as follows

$$\partial_{g^2}\Gamma = \Delta_g \cdot \Gamma \quad (27)$$

$$f_\eta(\xi)\mathcal{N}_A\Gamma \quad (28)$$
$$+\chi f'_\eta\left(\int Tr\left(\rho A - \bar{c}\frac{\delta\Gamma}{\delta B}\right) + 2\xi\partial_\chi\Gamma\right) = \Delta_\eta \cdot \Gamma$$

$$f_\sigma(\xi)\mathcal{N}_c\Gamma - \chi f'_\sigma \int Tr(\sigma c) = \Delta_\sigma \cdot \Gamma \quad (29)$$

$$f_\varphi(\xi)\mathcal{N}_\varphi\Gamma + \chi f'_\varphi \int Tr\underline{Y}\varphi = \Delta_\varphi \cdot \Gamma \quad (30)$$

$$\mathcal{N}_A \equiv N_A - N_\rho - N_B - N_{\bar{c}} + 2(\xi\partial_\xi + \chi\partial_\chi)$$
$$\mathcal{N}_c \equiv N_c - N_\sigma$$
$$\mathcal{N}_\varphi \equiv N_\varphi - N_Y$$
$$N_\psi \equiv \int \psi\frac{\delta}{\delta\psi} \quad \psi = A, \rho, \ldots$$

represent the basis of BRS symmetric differential operators and associated symmetric insertions. Their symmetry is expressed by

$$(\mathcal{S}_\Gamma + \chi\partial_\xi)\Delta_X \cdot \Gamma = 0 \quad X = g, \eta, \sigma, \varphi \quad (31)$$

Since (33) involves the variation of $\xi$, the appearance of the arbitrary function $f_X(\xi)$ $X = \eta, \sigma, \varphi$ of $\xi$, but no such function in front of $\partial_{g^2}$, is not accidental, but marks a significant distinction between these operators. It becomes relevant e.g. when we derive the RG-equ. The RG-equ. expresses the response of the theory to the variation of the normalization point $\kappa$. Since this is a symmetric operation as seen from

$$(\mathcal{S}_\Gamma + \chi\partial_\xi)(\kappa\partial_\kappa\Gamma) = 0, \quad (32)$$

the operator $\kappa\partial_\kappa$ can be expanded in the above basis with $\xi$-independent coefficients:

$$(\kappa\partial_\kappa + \tilde{\beta}_g\partial_{g^2} - \tilde{\gamma}_c f_\sigma \mathcal{N}_c - \tilde{\gamma}_A f_\eta \mathcal{N}_A$$
$$-\tilde{\gamma}_\varphi f_\varphi \mathcal{N}_\varphi)\Gamma = 0 \quad (33)$$

Hence the $\beta$-function of the RG-equ. appears as $\xi$-independent quantity, whereas the anomalous dimensions $\tilde{\gamma}f$ depend in general on $\xi$ due to the functions $f_X(\xi)$ present in (30-32).

The CS-equ. expresses the response of the theory to the scaling of all parameters of the model having mass dimensions, i.e. $\kappa$ and the physical masses $\underline{m}$. In addition to the hard differential operators (29) – (32) one has to construct soft BRS invariant mass insertions $\Delta_{\underline{m}}$ for all gauge invariant mass terms of the matter fields. Then one can derive as the CS-equ.

$$\mathcal{C}\Gamma = \alpha_{\underline{m}}\Delta_{\underline{m}} \cdot \Gamma \quad (34)$$
$$\mathcal{C} \equiv \kappa\partial_\kappa + \underline{m}\partial_{\underline{m}}$$
$$+\beta_g\partial_{g^2} - \gamma_c f_\sigma \mathcal{N}_c - \gamma_A f_\eta \mathcal{N}_A - \gamma_\varphi f_\varphi \mathcal{N}_\varphi \quad (35)$$

Again, the $\beta$-fct. is $\xi$-independent, whereas the anomalous dimensions in general depend on it.

### 2.4.2. Rigid gauge invariance spontaneously broken

For theories with spontaneously broken symmetry the construction of the CS- and RG-operators is much more involved [7,8], already without gauge fields. For gauge theories we quote here the result for the Abelian Higgs model. (The detailed derivation will be presented elsewhere [9].) The BRS invariant classical action reads

$$\Gamma_{inv} = \quad (36)$$
$$\int\left(-\frac{1}{4}F^{\mu\nu}F_{\mu\nu} + (D\phi)^*D\phi + \mu^2\phi^*\phi - \lambda(\phi^*\phi)^2\right)$$

where

$$\phi \equiv \frac{1}{\sqrt{2}}(\varphi_1 + v + i\varphi_2)$$
$$D\phi \equiv (\partial_\mu - ieA_\mu)\phi \quad (37)$$
$$F_{\mu\nu} \equiv \partial_\mu A_\nu - \partial_\nu A_\mu$$

and the BRS transformations are given by

$$s\varphi_1 = -ec\varphi_2$$

$$s\varphi_2 = ec(\varphi_1 + v)$$
$$sA_\mu = \partial_\mu c$$
$$sc = 0. \qquad (38)$$

The unphysical parameters $\mu, \lambda, v$ have to be expressed in terms of the physical parameters $m$ (vector mass), $m_H$ (Higgs mass), $e$ (charge) which are to be introduced by normalization conditions. Since we will have a particle interpretation only if the vacuum expectation values of the fields vanish and the particle poles are fixed with proper residue, we impose

$$<\varphi_1> \;=\; 0 \qquad (39)$$
$$\Gamma_{\varphi_1\varphi_1}(p^2 = m_H^2) \;=\; 0 \qquad (40)$$
$$\Gamma^T(p^2 = m^2) \;=\; 0 \qquad (41)$$

for $\Gamma_{\mu\nu} = -\left(\eta_{\mu\nu} - \frac{p_\mu p_\nu}{p^2}\right)\Gamma^T - \frac{p_\mu p_\nu}{p^2}\Gamma^L$.
For the residues

$$\partial_{p^2}\Gamma_{\varphi_1\varphi_1}(p^2 = \kappa^2) \;=\; 1 \qquad (42)$$
$$\partial_{p^2}\Gamma^T(p^2 = \kappa^2) \;=\; 1 \qquad (43)$$

And for the coupling

$$\partial_p \Gamma_{A\varphi_1\varphi_1}\Big|_{p_N} = e \qquad (44)$$

($p_N = p_N(\kappa)$ is a normalization point.) (The expectation value of $\varphi_2$ vanishes because it is odd under charge conjugation which we impose as discrete symmetry, that of $A_\mu$ because of Lorentz invariance.)

For the quantization one has to proceed as indicated above, namely one has to start from a gauge condition. In order to have mass degeneracy in the unphysical sector a 't Hooft type gauge fixing is appropriate (and two gauge parameter doublets):

$$\frac{\delta\Gamma}{\delta B} = \xi_1 B + \partial A + \xi_2 m\varphi_2 + \tfrac{1}{2}\chi_1 \bar{c}. \qquad (45)$$

Since now the BRS variation of the vector field is linear and that of $c$ vanishes one does not need the external fields $\rho_\mu, \sigma$ as in the non-abelian case. One only has to introduce external fields $Y_{1,2}$ coupled to the variations $s\varphi_{1,2}$ resp. In the classical approximation $\Gamma$ is therefore given by

$$\Gamma_{cl} = \Gamma_{inv} + \int(Y_1 s\varphi_1 + Y_2 s\varphi_2) \qquad (46)$$
$$+ s\int(\tfrac{1}{2}\xi_1 \bar{c}B + \bar{c}(\partial A + \xi_2 m\varphi_2))$$

The ST-identity has now the form

$$\mathcal{S}(\Gamma) \equiv \int\left(\partial c\frac{\delta\Gamma}{\delta A} + B\frac{\delta\Gamma}{\delta \bar{c}} + \frac{\delta\Gamma}{\delta \underline{Y}}\frac{\delta\Gamma}{\delta \underline{\varphi}}\right)$$
$$+ \underline{\chi}\partial_{\underline{\xi}}\Gamma = 0 \qquad (47)$$

Its validity requires as a necessary condition the ghost equ. of motion

$$\frac{\delta\Gamma}{\delta \bar{c}} + \xi_2 m\frac{\delta\Gamma}{\delta Y_2} = -\Box c - \tfrac{1}{2}\chi_1 B - \chi_2 m\varphi_2 \qquad (48)$$

to hold.

The ST identity (49) can be proved analogously to [10] with extended transformation laws. Together with charge conjugation invariance and with the normalization conditions indicated above it defines the theory.

What concerns the construction of symmetrical differential operators we have to make now a distinction which was not very important in the rigid symmetry case but which becomes important here. Since the gauge condition depends now on a mass parameter, we have to distinguish amongst differential operators $\nabla$ which are BRS symmetric, i.e. satisfy

$$(\mathcal{S}_\Gamma + \underline{\chi}\partial_{\underline{\xi}})(\nabla\Gamma) = 0 \qquad (49)$$

with

$$\mathcal{S}_\Gamma \equiv \int\left(\partial c\frac{\delta}{\delta A} + B\frac{\delta}{\delta \bar{c}} + \frac{\delta\Gamma}{\delta Y_1}\frac{\delta}{\delta\varphi_1} + \frac{\delta\Gamma}{\delta\varphi_1}\frac{\delta}{\delta Y_1}\right.$$
$$\left. + \frac{\delta\Gamma}{\delta Y_2}\frac{\delta}{\delta\varphi_2} + \frac{\delta\Gamma}{\delta\varphi_2}\frac{\delta}{\delta Y_2}\right) \qquad (50)$$

those which commute with the gauge condition and those which do *not*. Without going into more detail let us here only mention that a basis of the BRS symmetric differential operators which commute with the gauge condition is spanned by

$$e\partial_e,\; m_H\partial_{m_H},\; m\partial_m - (\xi_2\partial_{\xi_2} + \chi_2\partial_{\chi_2}) \equiv m\tilde{\partial}_m \quad (51)$$

$$\mathcal{N}_A \equiv \int\left(A^\mu\frac{\delta}{\delta A^\mu} - B\frac{\delta}{\delta B} + c\frac{\delta}{\delta c} - \bar{c}\frac{\delta}{\delta \bar{c}}\right)$$
$$+ 2D_1 + D_2 \qquad (52)$$
$$\mathcal{N}_\varphi \equiv \int\left(\varphi_1\frac{\delta}{\delta\varphi_1} + \varphi_2\frac{\delta}{\delta\varphi_2} - Y_1\frac{\delta}{\delta Y_1} - Y_2\frac{\delta}{\delta Y_2}\right)$$
$$- D_2 \qquad (53)$$

In the basis of BRS symmetric differential operators which do not commute with the gauge condition the above $\mathcal{N}_A$ is split into

$$\mathcal{N}_A \equiv \int \left( A^\mu \frac{\delta}{\delta A^\mu} + c \frac{\delta}{\delta c} \right) \qquad (54)$$

$$\mathcal{N}_B \equiv \int \left( B \frac{\delta}{\delta B} + \bar{c} \frac{\delta}{\delta \bar{c}} \right) \qquad (55)$$

$$D_i \equiv \xi_i \partial_{\xi_i} + \chi_i \partial_{\chi_i} \qquad i = 1, 2 \text{ no sum} \qquad (56)$$

The operator $\mathcal{N}_\varphi$ is changed into the form (32), i.e. becomes $\xi$-dependent.

Varying the normalization point is a BRS symmetric operation which respects the gauge condition, hence $\kappa \partial_\kappa \Gamma$ can be expanded in the smaller basis and the RG-equ. reads

$$(\kappa \partial_\kappa + \tilde{\beta}_e e \partial_e - \tilde{\gamma}_A \mathcal{N}_A - \tilde{\gamma}_\varphi \mathcal{N}_\varphi) \Gamma = 0 \qquad (57)$$

Since we have normalized physically (i.e. pole normalization), the coefficients of the operators differentiating with respect to mass parameters vanish. It is noteworthy that all coefficient functions are $\xi$-independent.

The dilatation operator $m\partial_m + m_H \partial_{m_H} + \kappa \partial_\kappa$ does not commute with the gauge condition hence its hard part has to be expanded in the larger basis of only BRS symmetric operators. The CS-equ. results only after construction of a basis for the soft BRS symmetric insertions. This we shall skip here and only mention that it is possible (with the help of additional external fields $q_o, q_1$). We thus arrive for the CS-equ. at

$$\mathcal{C}\Gamma = \alpha m \int \left( \frac{\delta}{\delta \varphi_1} + \hat{\alpha}_o \frac{\delta}{\delta q_o} + \hat{\alpha}_1 \frac{\delta}{\delta q_1} \right) \Gamma^q \bigg|_{\tilde{q}=0} \qquad (58)$$

$$\begin{aligned} \mathcal{C} \equiv\ & m\partial_m + m_H \partial_{m_H} + \kappa \partial_\kappa \\ & + \beta_e e \partial_e + \beta_m m \tilde{\partial}_m + \beta_{m_H} m_H \partial_{m_H} \\ & - \gamma_A \mathcal{N}_A - \gamma_B \mathcal{N}_B - \gamma_\varphi f_\varphi \mathcal{N}_\varphi \\ & + \sum_{i=1}^{2} \gamma_i D_i \end{aligned} \qquad (59)$$

This form shows that from all coefficient functions only the anomalous dimensions of the matter fields depend on $\xi$, whereas all others turn out to be $\xi$-independent. Most remarkable is the fact that in the CS-operator $\beta$-functions with respect to the physical masses show up. Beginning with two loops they will affect the theory.

## 3. Conclusions

BRS invariance enlarged by doublets of gauge parameters varying into Grassmann numbers permits a simple control of gauge parameter dependence in gauge theories. As gauge independent quantities we displayed explicitly: the coefficient of the chiral anomaly, elements of the physical $S$-matrix and $\beta$-functions. But also out of the vector two-point-function in Yang-Mills theory one can construct a gauge independent quantity. The case of gauge theories with spontaneously broken rigid invariance requires and deserves further study. Here we have presented only the example of the Abelian Higgs model.

Once corresponding results have been obtained for the standard model one can study in this context other quantities of phenomenological interest such as the $S, T, U$-parameters.